\documentclass{article}

\PassOptionsToPackage{numbers}{natbib}
\usepackage[preprint]{neurips_2021}

\usepackage[utf8]{inputenc} 
\usepackage[T1]{fontenc}    
\usepackage{booktabs}       
\usepackage{amsfonts}       
\usepackage{nicefrac}       
\usepackage{microtype}      
\usepackage{amsmath}
\usepackage{graphicx}
\usepackage{tipa}
\usepackage{multirow}
\usepackage{longtable}

\usepackage{subcaption}
\usepackage{color}
\usepackage[table]{xcolor}
\usepackage{threeparttable}
\usepackage{diagbox}
\usepackage{enumitem}
\usepackage{tablefootnote}
\usepackage{hyperref}
\usepackage{url}            
\usepackage{xurl}
\usepackage{breakurl}

\usepackage[edges]{forest}
\usepackage{tikz}
\usepackage{tikz-qtree}
\usepackage{svg}
\svgpath{{img/}}
\usetikzlibrary{fadings}
\usetikzlibrary{mindmap,trees}
\usetikzlibrary{arrows,automata,shapes,positioning,shadows,trees}
\usetikzlibrary{shapes,snakes,shadows}
\usetikzlibrary{shapes.arrows}
\usetikzlibrary{calc,shapes, positioning}
\usetikzlibrary{decorations.text}
\usetikzlibrary{matrix,chains,positioning,decorations.pathreplacing,arrows}
\usetikzlibrary{bayesnet}
\usetikzlibrary{shadows.blur}
\usetikzlibrary{shapes.geometric}

\tikzstyle{edge}=[-latex',draw=black!90,shorten <=1pt,shorten >=1pt]
\tikzstyle{redge}=[latex'-,draw=black!90,shorten <=1pt,shorten >=1pt]
\tikzstyle{dedge}=[latex'-latex',draw=black!90,shorten <=1pt,shorten >=1pt]
\tikzstyle{block}=[draw, text width=5em,align=center,shape=rectangle, rounded corners, , align=center]
\tikzstyle{nobox}=[align=center]

\definecolor{emb}{RGB}{209,228,252}
\definecolor{hidden-blue}{RGB}{194,232,247}
\definecolor{hidden-orange}{RGB}{243,202,120}
\definecolor{hidden-yellow}{RGB}{242,244,193}
\definecolor{output-purple}{RGB}{219,203,231}
\definecolor{output-green}{RGB}{204,231,207}
\definecolor{hiddendraw}{RGB}{205, 44, 36}

\tikzstyle{mybox}=[
    rectangle,
    draw=hiddendraw,
    rounded corners,
    text opacity=1,
    minimum height=1.5em,
    minimum width=5em,
    inner sep=2pt,
    align=center,
    fill opacity=.2,
    ]

\tikzstyle{emb-purple}=[
    rectangle,
    draw=output-purple!50!purple,
    fill=output-purple,
    text opacity=1,
    minimum height=1.5em,
    minimum width=1.5em,
    inner sep=0pt,
    align=center,
    fill opacity=.9,
    ]

\tikzstyle{emb-blue}=[
    rectangle,
    draw=emb!50!blue,
    fill=emb,
    text opacity=1,
    minimum height=1.5em,
    minimum width=1.5em,
    inner sep=0pt,
    align=center,
    fill opacity=.9,
]

\definecolor{colone}{RGB}{178, 34, 34}
\definecolor{coltwo}{RGB}{106, 90, 205}
\definecolor{colthree}{RGB}{255, 250, 205}
\definecolor{colfour}{RGB}{0, 139, 69}
\definecolor{colfive}{RGB}{245,238,197}
\definecolor{colsix}{RGB}{243,235,179}
\definecolor{colseven}{RGB}{241,231,163}
\title{A Survey about application based on RF signals}

\author{
Jiaren Xiao\\
\texttt{212020081200026@stu.xhu.edu.cn} \\
Xihua University \\
}

\begin{document}
\maketitle

\begin{abstract}
    With the popularity of commercial wireless devices, the research based on RF signal has developed rapidly in the past decade. RF signals have great advantages in environmental perception. Wireless signals are transmitted through the transmitter. With the propagation of these signals in the medium, they are reflected from different objects and people in space to the receiver. In this process, rich environmental perception information is carried. These ubiquitous perceptual information has rich application significance and bright prospects in the fields of indoor positioning, motion perception, behavior recognition, human-computer interaction, fall detection, health monitoring, smart home, search and rescue and so on. This paper will discuss and analyze the current research results by combing some existing literature and the research status of relevant researchers in relevant fields at this stage, so as to understand the future development trend of RF signal based applications.
\end{abstract}

\section{introduction}
\label{inrtoduction}
The application of wireless signal is mainly divided into two directions: WiFi Based and radar based. Compared with WiFi, radar has stronger anti-environmental interference and can carry more environmental information and object information. 
This paper mainly summarizes the RF applications based on radar. 
According to the working mode, the radar can be divided into pulse radar\cite{arbabian201394, heunisch2019millimeter, park2016ir, kim2017hand} and continuous wave radar\cite{adib2015multi,sun2020real,lien2016soli,pramudita2020time}. 
Pulse radar intermittently transmits rectangular pulse periodic signals and receives reflected echo signals in the transmission gap to perceive the environment, but pulse radar has a blind area for short-range detection.
Continuous wave radar can be further subdivided into single frequency continuous wave (SFCM) and frequency modulated continuous wave (FMCW)\cite{adib2015multi}. 
It can sense the environment by transmitting continuous wave and receiving the reflected echo signal at the same time, but countinuous wave radar has the problems of signal leakage and background interference. 
Single frequency continuous wave can only be used for velocity measurement and cannot be used for ranging. FMCW can be used not only for velocity measurement and ranging, but also for distinguishing moving targets. 
Appropriate radar signal technology can be flexibly selected according to specific research.

Indoor location is the basis of RF research, and the premise of many studies is the need for location information. 
In 2013, Fadel Adib used a 3-antenna MIMO radio with low bandwidth and low power consumption\cite{adib2013see} to determine the number of people in closed rooms and their relative positions through the wall, and showed that it has great limitations to track the human body through the RF signal generated by human motion. 
In 2014, an indoor 3D positioning system WiTrack\cite{adib20143d} was proposed to realize the accurate positioning of indoor human body using RF signal, but this positioning system can only locate people in motion. 
In 2015, based on the improvement of WiTrack, WiTrack 2.0\cite{adib2015multi} was proposed, in which the accurate positioning of multi person scene and static human body was realized. 
In the same year, a through wall human body contour capture system RF capture\cite{adib2015capturing} was proposed to track the 3D position of a person's limbs and body parts, generate a coarse-grained human body contour, and distinguish different users through the contour information, which laid a foundation for subsequent human body posture estimation. 
Subsequently, with the rapid development of perception technology and deep learning based on RF signal, a series of applications such as sleep detection\cite{yue2020bodycompass,zhao2017learning}, fall detection\cite{alnaeb2019detection}, gesture recognition\cite{sun2020real,lien2016soli,arbabian201394,heunisch2019millimeter,park2016ir,kim2017hand,skaria2019hand,miller2020radsense,pramudita2020time}, 
radar imaging\cite{bocca2013multiple,adib2015multi,adib2015capturing,zhao2018through},
physiological feature monitoring\cite{yue2020bodycompass,yue2018extracting,liu2018non,chen2020respiration,adib2015smart} , gait recognition\cite{kim2014human,vandersmissen2018indoor} , track tracking\cite{hsu2017extracting,hsu2019enabling} have been realized. It is widely used in video games, intelligent video surveillance, smart home, human-computer interaction, security and other fields, and has broad application prospects and economic value.

Environment sensing technology based on wireless signal has many unique advantages. 
Firstly, without wearing any sensors, behavior perception based on wireless signals is realized by detecting the RF signal characteristics reflected by the human body. Secondly, it has good obstacle crossing ability and realizes non line of sight perception. 
Third, radar signal belongs to electromagnetic wave and is less affected by external conditions such as light, temperature and humidity. 
Fourth, it has strong scalability. With the continuous development of radar technology, the bandwidth and working frequency of radar are increasing, and the number of antennas is also increasing, which lays a good foundation for further improving the sensing accuracy.

In this review, we mainly review the research work based on RF in recent years, and the main contributions are as follows:
\begin{itemize}
    \item In section~\ref{Background}, we introduce the ranging principle, angle measurement principle and velocity measurement principle based on radar signal.
    \item In section~\ref{Data preprocessing}, we describe the flow and method of radar signal preprocessing.
    \item In section~\ref{RF application}, we introduce applications based on RF signals. It mainly includes indoor positioning, gesture recognition, physiological signal monitoring, human activity recognition and pose estimation.
\end{itemize}

\tikzstyle{leaf}=[mybox,minimum height=1.2em,
fill=hidden-orange!50, text width=5em,  text=black,align=left,font=\footnotesize,
inner xsep=4pt,
inner ysep=1pt,
]

\begin{figure*}[thp]
 \centering
\begin{forest}
  forked edges,
  for tree={
  grow=east,
  reversed=true,  
  anchor=base west,
  parent anchor=east,
  child anchor=west,
  base=left,
  font=\normalsize,
  rectangle,
  draw=hiddendraw,
  rounded corners,
  align=left,
  minimum width=2.5em,
  inner xsep=4pt,
  inner ysep=0pt,
  },
    [Survey
        [Sec.~\ref{Background}: Background
            [Sec.~\ref{Ranging principle}: Ranging principle]
            [Sec.~\ref{Angle measurement principle}: Angle measurement principle]
            [Sec.~\ref{Velocity measurement principle}: Velocity measurement principle]
        ]
        [Sec.~\ref{Data preprocessing}: Data preprocessing
            [Sec.~\ref{Eliminating flash effect}: Eliminating flash effect]
            [Sec.~\ref{Eliminate multi-path effect}: Eliminate multi-path effect]
            [Sec.~\ref{Elimination of near-far effect}: Elimination of near-far effect]
            [Sec.~\ref{Filter denoising}: Filter denoising]
        ]
        [Sec.~\ref{RF application}: RF \\application
            [Sec.~\ref{Indoor positioning}: Indoor positioning
                [Sec.~\ref{Base on RSSI}: Base on RSSI
                ]
                [Sec.~\ref{Base on AOA}: Base on AOA
                ]
                [Sec.~\ref{Base on TOF}: Base on TOF
                ]
            ]
            [Sec.~\ref{Gesture recognition and physiological sign monitoring }: Gesture recognition and \\physiological sign monitoring
                [Sec.~\ref{Gesture recognition}: Gesture \\recognition
                ]
                [Sec.~\ref{Physiological sign monitoring}: Physiological \\sign monitoring
                ]
            ]
            [Sec.~\ref{Human activity recognition}: Human \\activity recognition
                [Sec.~\ref{Identity recognition}: Identity recognition
                ]
                [Sec.~\ref{Activity recognition}: Activity recognition
                ]
            ]
            [Sec.~\ref{Human pose estimation}: Human pose estimation
            ]
        ]
        [Sec.~\ref{Conclusion}: Conclusion
        ]
    ]
\end{forest}
\caption{Organization of this paper.}
\label{org_survey_paper}
\end{figure*}
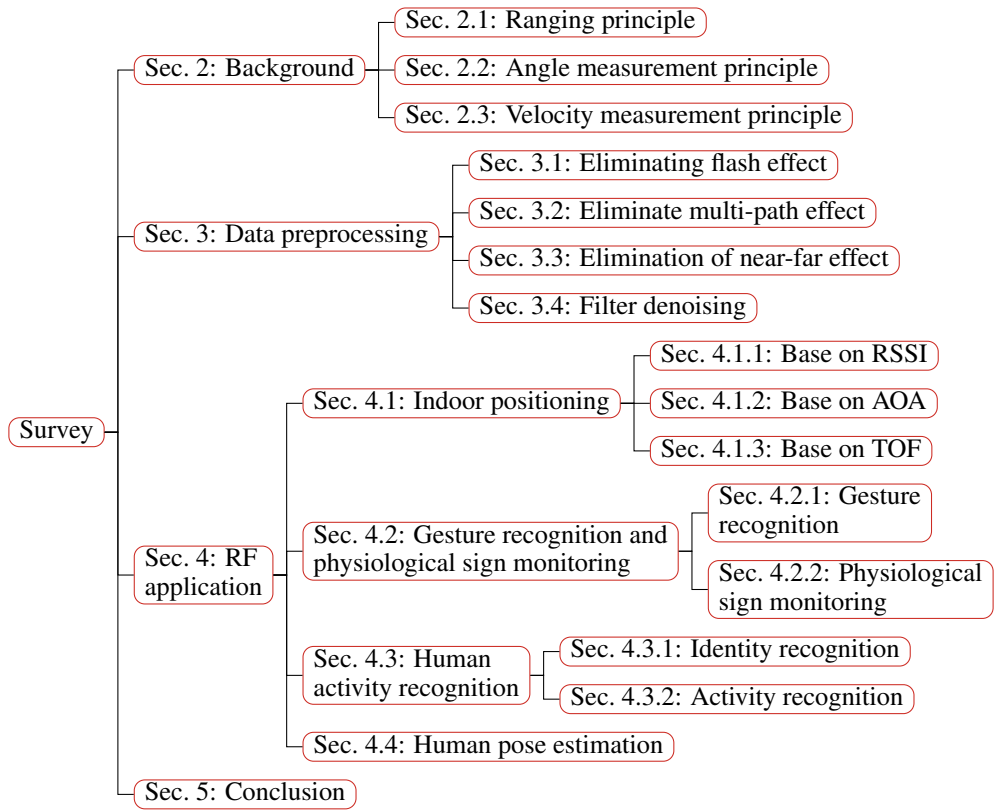

\section{Background}
\label{Background}
Radar transmits electromagnetic energy pulses through the transmitter, and RF energy propagates to and reflects from the reflected object in the medium. A small part of the reflected energy is transmitted to the receiver of the radar device, which uses this echo to determine the direction, distance and speed of the reflected object. Assuming that the FM signal is used as the transmission signal, the transmission signal can be expressed as follows:
\begin{equation}
\label{eq1}
    S(t) = A_te^{j2\pi(f_ct + \frac{1}{2}kt^2)}
\end{equation}
Where, $A_t$ represents the amplitude at time t, $f_c$ represents the radar carrier frequency, and $k$ represents the frequency modulation slope.
The reflected signal propagates in space and is reflected by the target object to the receiving antenna, so received signal can be expressed as follow:
\begin{equation}
\label{eq2}
    R(t) = A_te^{j2\pi(f_c(t - \tau) + \frac{1}{2}k(t - \tau)^2)}
\end{equation}
Where, $\tau = \frac{2R}{C}$, represents the instantaneous time delay of echo, $A_t$ represents the amplitude at time t, $f_c$ represents the radar carrier frequency, and $k$ represents the frequency modulation slope.

\subsection{Ranging principle}
\label{Ranging principle}
The radar transmits a radio pulse signal with very high power. The pulse signal is concentrated in one direction through the directivity of the antenna, and a speed of light propagates in a given direction. If there is a target object in this direction, part of the pulse energy will be reflected back to the radar. The radar receives this energy signal and estimates the target object distance R. Specific formula as follows:
\begin{equation}
\label{eq3}
    R = \frac{C\cdot t}{2}
\end{equation}
Where $C$ represents the speed of light, $t$ represents the propagation time of the signal in the medium, and $R$ represents the oblique departure of the target object (line of sight distance). 

In addition, the distance resolution is shown in follow:
\begin{equation}
\label{eq4}
    R_{res} = \frac{C}{2B}
\end{equation}
Where $C$ represents the speed of light and $B$ represents the bandwidth. It can be seen from the equation \ref{eq4} that the greater the bandwidth, the better the distance resolution.

\subsection{Angle measurement principle}
\label{Angle measurement principle}
\begin{figure}[htbp]
    \centering
    \includegraphics[height = 5cm]{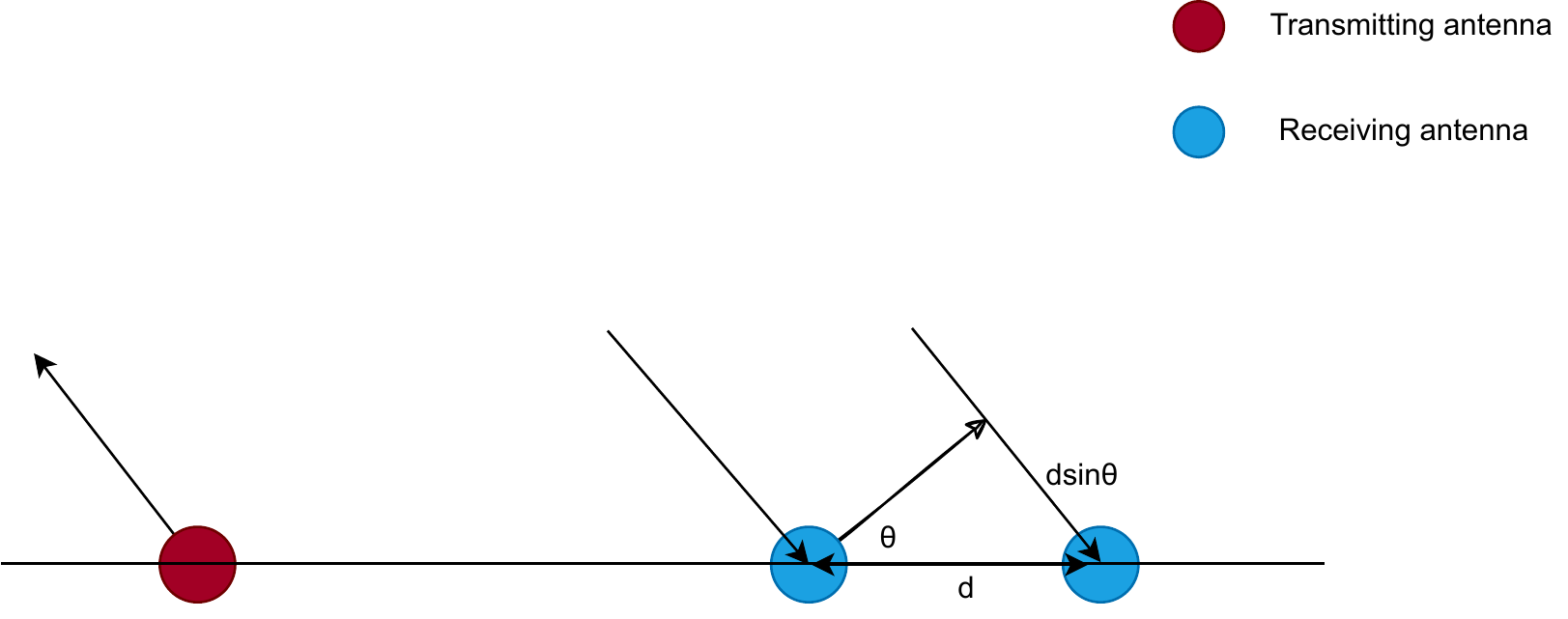} 
    \caption{\textbf{A sample example of angle measuring. }\footnotesize{The transmitting antenna transmits the signal and reflects it back after touching the target object. The azimuth of the target can be calculated through the phase difference of the signals received by different receiving antennas.}}
    \label{fig1}
\end{figure}

The angle of the target is determined by the directivity of the antenna. By measuring the direction of the antenna when receiving the echo, the azimuth $\theta$ and elevation angle $\phi$ from the radar to the target object can be determined. The simplest angle measurement principle is shown in Figure \ref{fig1}. The phase difference of two receiving antenna signals caused by distance difference is expressed as:
\begin{equation}
\label{eq5}
    \omega = \frac{2\pi}{\lambda}d\sin\theta
\end{equation}
So we can get
\begin{equation}
    \label{eq6}
    \theta = \sin^{-1}(\frac{\omega\lambda}{2\pi d})
\end{equation}
In this way, the angle of antenna signal transmission is obtained, where $d$ represents the distance between receiving antennas, $\omega$ represents the phase difference between two antennas, and $\lambda$ represents the wavelength.

In addition, the angular resolution is approximately as follow:
\begin{equation}
    \label{eq7}
    \theta_{res} \approx \frac{2}{N}(rad)
\end{equation}
Where $N$ represents the virtual antenna array element. According to the equation \ref{eq7}, MIMO radar technology is used to form an array element of virtual receiving antenna with the product of the number of transmitting antennas and the number of receiving antennas. By adding $N$, the angular resolution of the radar can be simply and effectively improved.

\subsection{Velocity measurement principle}
\label{Velocity measurement principle}
For a moving target, there is a frequency $\Delta f$ difference between the transmitted signal and the received signal, which is called Doppler frequency, so the target speed $V$ can compute as follow:
\begin{equation}
    \label{eq8}
    V = \frac{f_d\cdot \lambda}{2} = \frac{\lambda \omega}{4\pi T_c}
\end{equation}
Where $f_d$ represents the Doppler frequency, $\lambda$ represents the wavelength, $\omega$ represents the phase difference and $T_c$ represents a chirp period.
In addition, it can be deduced that the maximum measurable unambiguous speed of the radar as follow:
\begin{equation}
    \label{eq9}
    V_{max} = \frac{\lambda}{4T_c}
\end{equation}
From equation \ref{eq9}, it can be seen that the constraint of the maximum unambiguous speed is the phase difference, which between positive and negative 180rad, and the larger measurable unambiguous speed needs more dense chirp.
In addition, the velocity resolution can be compute as the follow equation:
\begin{equation}
    \label{eq10}
    V_{res} = \frac{\lambda}{2NT_c}
\end{equation}
Where $\lambda$ represents the wavelength and $NT_c$ represents the time of one frame. The shorter the wavelength, the longer the time of one frame, and the higher the speed resolution.

\section{Data preprocessing}
\label{Data preprocessing}
\subsection{Eliminating flash effect}
\label{Eliminating flash effect}
When the equipment collects the human body reflected signal behind the wall, because the RF signal will penetrate the Wall twice, the power of the signal itself will drop by several orders of magnitude, and the reflected signal from the wall will be much larger than the reflected signal from the human body. This strong reflection from the wall will drown the analog-to-digital converter of the receiver, Preventing it from recording small changes in the reflected signal from the real human body is called the "flash effect". To solve this problem, either increase the signal bandwidth so that it can record small changes from behind the wall, isolate the reflection of the wall from the reflection of the object behind the wall, or install an additional receiving antenna behind the wall to synchronize the time of the received signal. In\cite{adib2013see}, a MIMO interference zeroing technology is proposed to eliminate the flash effect. The whole process is divided into three stages: initialization zeroing, power boost and iterative zeroing. Firstly, two known signals are transmitted on the two transmitting antennas in turn, and then the channels between the two transmitting antennas and the receiving antenna are estimated with the signals received by the receiving end, and then the two channels are used to calculate the ratio. Through this ratio, the state in the joint channel can be calculated. If no object moves, the channel is empty, which eliminates the reflected signals from all static objects.

\subsection{Eliminate multi-path effect}
\label{Eliminate multi-path effect}
Multi-path effect\cite{adib2015multi,adib20143d} consists of two parts. One part is the static background emission in the environment, which is called static multi-path; The other part is the signal reflected by the human body to the static background, and then reflected by the static background to the receiving antenna, which is called dynamic multi-path. 
This multipath effect is reflected in radar imaging, which will produce multiple spots. This spot will not prevent the system from detecting humans, but will hinder human reflection and multipath reflection, thus limiting the ability to locate humans in the real environment.
Since the TOF time generated by static objects in the environment does not change with time, based on this feature, the interference of static multi-path effect is eliminated by subtracting continuous RF frames. The elimination of dynamic multi-path is more complex. Because the dynamic multi-path is caused by the indirect reflection of the human body, and the TOF will change with time, the idea of dynamic multi-path elimination is that the path of the receiving antenna receiving the direct reflection from the human body at any time point is shorter than any indirect reflection path. Since distance and TOF are directly related, and TOF is related to frequency, distance and frequency are directly related. Based on this idea, the reflection through the shortest path is tracked by tracking the bottom contour (a local maximum closest to the device) in all strong reflection surfaces, so as to eliminate the dynamic multi-path.

\subsection{Elimination of near-far effect}
\label{Elimination of near-far effect}
The near-far effect is a major challenge in multi-person positioning, which means that the intensity of the signal reflected by the human body close to the equipment is much higher than that far away from the equipment, which will blur the signal from the distant person, resulting in inaccurate positioning and even undetected. In order to eliminate this near far effect, SSC\cite{adib2015multi} algorithm is proposed, which is divided into four steps: SSC detection, SSC remapping, SSC elimination and SSC iteration. Find the position of the strongest human body by superimposing the heat-maps of signals from different antennas at the same time, then remapping the position of this person into TOF, then set the TOF configuration of this point in all antenna pairs to zero, finally recalculate the heat map according to the new TOF configuration file, and then repeat the above steps until all people are found.

\subsection{Filter denoising}
\label{Filter denoising}
In addition to normal reflection signals, the collected signals are often mixed with random noise signals, which are caused by the stability of equipment level and weak anti-interference. These noise signals are not of interest to us and will affect the accuracy of the whole experiment. The existence of these random noises is inevitable. The influence of noise on experimental results can only be reduced by interpolation, deletion and filtering. The idea of interpolation is that the noise is accidental, and the normal signal value before and after the abnormal value is used to replace the abnormal value. Deletion is to discard the whole abnormal signal. Since the uniform motion of the human body is continuous and smooth, a filter can be used to smooth the convex outliers, For example, Butterworth filter, discrete wavelet filter, Kalman filter etc.

\section{RF application}
\label{RF application}
\subsection{Indoor positioning}
\label{Indoor positioning}
Indoor locating is an important field of RF application. The work of early researchers mainly focused on Simulation and modeling. In the last decade, researchers began to shift their attention to mobile humans, and even through wall positioning can be realized. These indoor positioning methods can be divided into equipment positioning and no equipment positioning. For device positioning, individuals need to wear RF signal transmitting devices, such as mobile phones, smart bracelets, etc. The equipment free positioning technology\cite{adib20143d,bocca2013multiple,sasakawa2016evaluation,adib2013see,adib2015multi} does not need to wear equipment on the body. It uses the reflected RF signal of the human body to track the human body positioning. Because this positioning technology completely depends on the weak and low-power RF reflection outside the human body, which is easily blocked by the indoor complex environment, such as metal objects, mirrors, walls, etc. The work of early researchers mainly focused on Simulation and modeling. In the last decade, researchers began to shift their attention to mobile humans. In terms of positioning principle, it can be divided into RSSI\cite{sun2020construction}, AOA\cite{adib2013see} and TOF\cite{adib20143d,adib2015multi}. The location technology based on RSSI mainly depends on the attenuation of the signal to determine the location. AOA determines the position according to the angle between the human body and different antennas. TOF determines the position by the time the signal propagates in space.

\subsubsection{Base on RSSI}
\label{Base on RSSI}
When the distance is close, the RSSI attenuation is faster, and when the distance is far, the RSSI attenuation is slower. The positioning technology based on RSSI estimates the signal strength value between the transmitting point and the receiving point according to this characteristic, and then converts the propagation loss into the distance between the receiving point and the transmitting point, and then calculates the location of the unknown node according to the relevant algorithm. In \cite{sun2020construction}, a positioning framework based on hybrid dual frequency RSSI is proposed. By setting multiple reference positioning points in advance, and then using a smart phone to collect and preprocess the dual RF RSSI, a dual RF RSSI fingerprint model is established to improve the positioning accuracy. This positioning technology based on RSSI has low hardware requirements, low power consumption and easy implementation. It is suitable for simple environment. Once the indoor environment is complex, the positioning accuracy is greatly reduced. Therefore, this positioning technology is generally suitable for small-scale and simple environment.

\subsubsection{Base on AOA}
\label{Base on AOA}
\begin{figure}[htbp]
    \centering
    \includegraphics[height = 5cm]{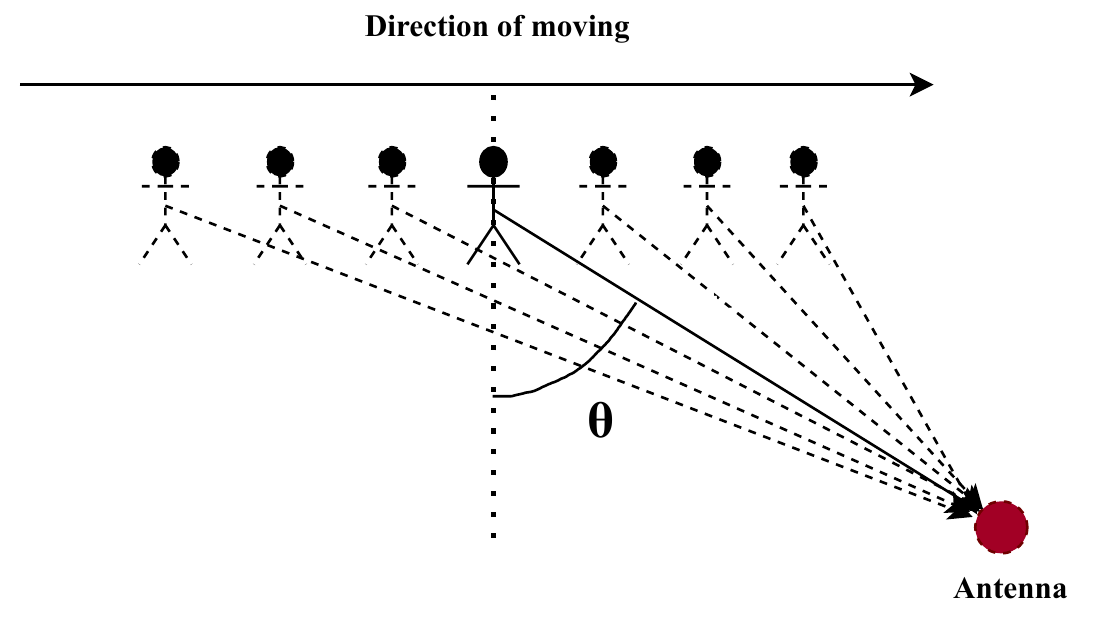} 
    \caption{\textbf{ISAR. }\footnotesize{The moving object itself emulates an antenna array as an inverse synthetic aperture. We leverages this principle to locate the moving target object.}}
    \label{fig2}
\end{figure}

In\cite{adib2013see}, a Wi-Vi system based on AOA is proposed. This method can even go through the wall to determine the number of people and their relative position in an enclosed space, and recognize the simple actions made by others behind the wall. Through tracking calculation with conventional antenna array $\Theta$ In contrast to locating stationary users, Wi-VI uses a receiving antenna and uses inverse synthetic aperture radar technology(show in  Figure \ref{fig2}) to compare the position of mobile users at different times as an antenna array for calculation $\Theta$, Locate and track dynamic users.

\subsubsection{Base on TOF}
\label{Base on TOF}
\begin{figure}[htbp]
    \centering
    \includegraphics[height = 5cm]{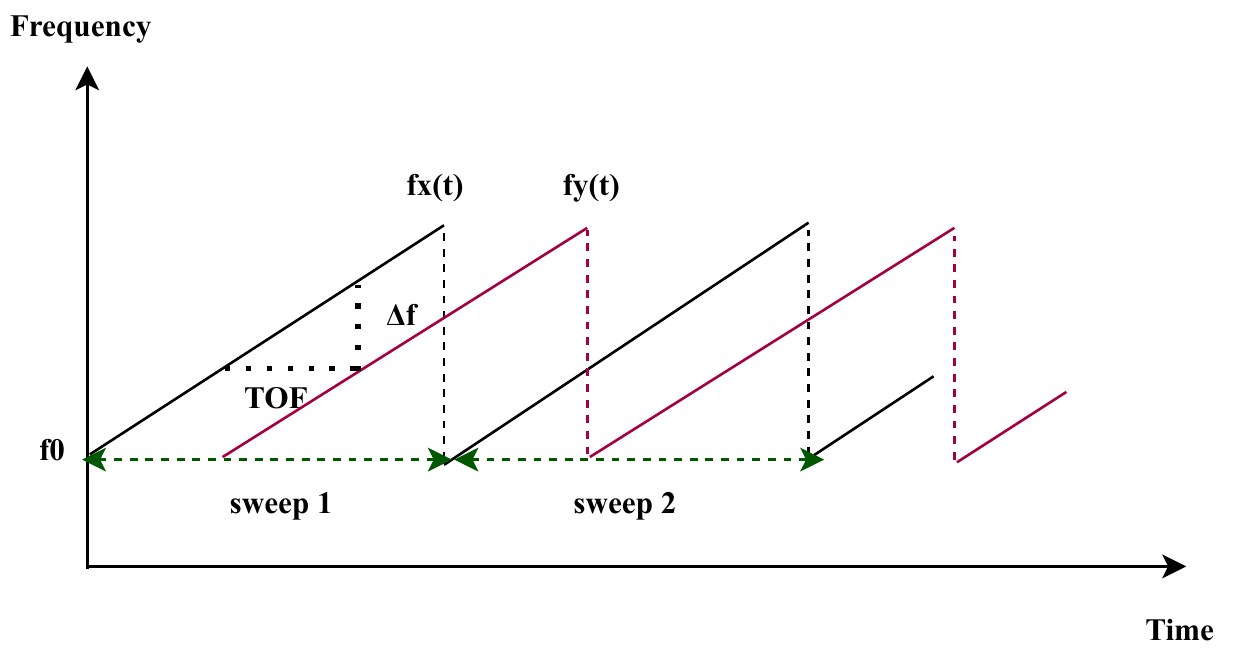} 
    \caption{\textbf{FMCW operation. }\footnotesize{For the carrier frequency sweep signal of the transmission period of the transmitting antenna, the received signal $f_y(t)$ has a time-shifted relative to the transmission signal $f_x(t)$, which is reflected as a frequency-shifted in the carrier frequency. TOF can be obtained by this frequency-shift $\delta f$}.}
    \label{fig3}
\end{figure}

\textbf{Acquisition of TOF: }
The measurement of TOF\cite{adib20143d,adib2015multi} is difficult. Because the RF signal propagates in space at the speed of light, the difference of tens of centimeters in space may be the reflection time difference of tens of microseconds in time dimension, and the measurement of tens of microseconds is difficult. The most direct way is to design a sub nanosecond high-speed digital to analog converter to sample the transmitted pulse signal and obtain the TOF time. However, this method is expensive and cost-effective. Therefore, a TOF estimation method based on FMCW is designed. As shown in the Figure \ref{fig3}, the change in time is transformed into the change in frequency by using the characteristics of FMCW. The carrier frequency of a series of chirp signals transmitted by FMCW changes linearly in the cycle time, showing a repeated frequency scanning operation in the whole time dimension. FMCW converts the time difference into the carrier frequency offset, which is easy to observe in the spectrum of the received signal.

\begin{figure}[htbp]
    \centering
    \includegraphics[height = 5cm]{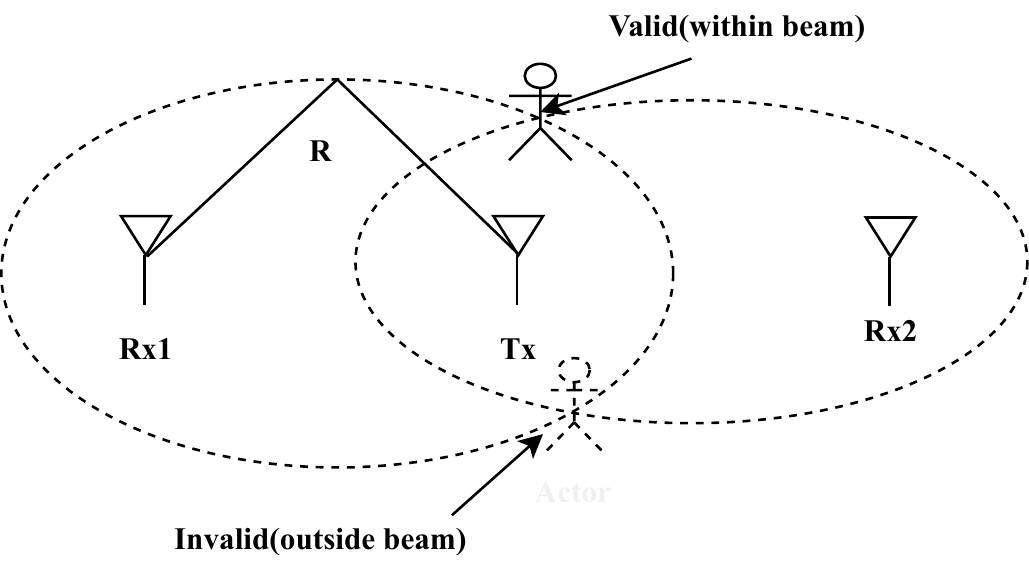} 
    \caption{\textbf{2D Localization. }\footnotesize{The TOF estimate from a receiving antenna defines an ellipse whose foci are the transmitting antenna and the receiving antenna. it can uniquely localize a person using the intersection of two ellipses. }}
    \label{fig5}
\end{figure}

\textbf{Single person positioning: }
The WiTrack\cite{adib20143d} system calculates the TOF through the frequency difference between the transmitted signal and the reflected signal, and then performs single person positioning in combination with the geometric position of the horizontal antenna. Through the estimated TOF calculation, a set of user positions is obtained. Mapping the set of positions into 2D space can obtain an ellipse. The ellipses obtained by two groups of different horizontal antennas will produce two groups of intersections in space, and then a group of intersection positions can be excluded according to the orientation of the antenna, so as to realize 2D positioning (as shown in the Figure \ref{fig5}). 3D positioning is to add a set of vertical antenna positions on the basis of two-dimensional.

\begin{figure}[htbp]
    \centering
    \includegraphics[height = 5cm]{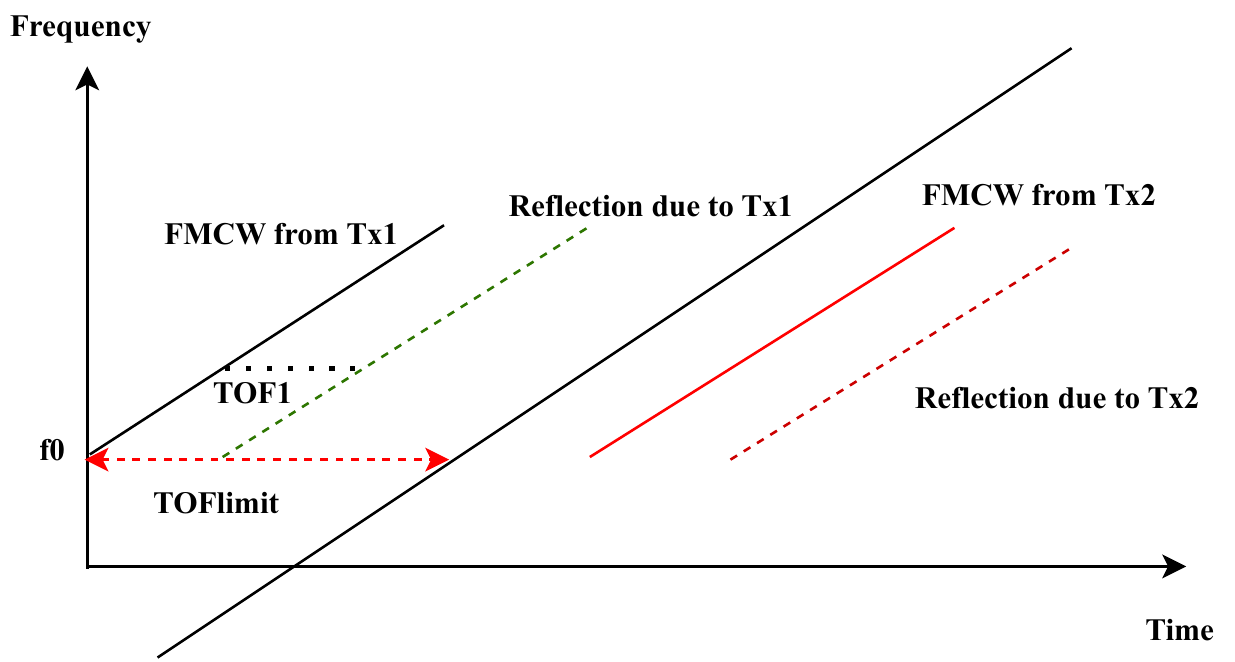} 
    \caption{\textbf{Multi-shift FMCW. }\footnotesize{Transmits FMCW signals from different transmitting antennas after inserting virtual delays $\tau$ between them. Each delay must be larger than the highest time-of-flight($TOF_limit$).}}
    \label{fig4}
\end{figure}

\textbf{Multi person positioning and static user positioning: }
WiTrack2.0\cite{adib2015multi} has been improved on the basis of WiTrack, which can not only locate multiple people, but also locate the static human body through the slight movement generated by human breathing. Previous positioning systems either needed dense and measured sensor grids to cover the whole space, or could not locate multiple users, which made them unsuitable for home environment. In addition, they require users to move constantly to detect their presence, requiring extensive prior calibration or training. WiTrack 2.0 solves these problems. It is a device free positioning system with multiple transmit and receive antennas. In the multi person positioning, the biggest challenge is how to eliminate the signal interference in the space medium and correctly match the TOF with the antenna. In multi person positioning, different people have different distances from the same antenna, and the same antenna has different distances from different people. As a result, multiple peaks will appear in the TOF profile of the received signal in the receiving antenna, and multiple TOF measurements will be obtained. If we can't correctly match different TOF measurements with the correct antenna pair, a set of wrong ellipses will be generated, which will locate people in the wrong position. In order to solve this challenge, it can be set a transmission interval $\tau$ to eliminate the signal interference in the space medium and $\tau > TOF_{limit}$ ($TOF_{limit}$ is the largest TOF in the environmental reflected signal), antennas transmit signals in different transmission $\tau$ successively at intervals(as show in Figure \ref{fig4}), so that different TOFs can be correctly matched with antenna pairs to accurately locate multiple users in the environment, Get the correct spatial location of the user. Static user positioning is different from non static user positioning. It uses long window subtraction for positioning and frames with long interval for static user positioning. The static user is positioned according to the small motion generated by inhalation and exhalation. Because the human respiratory motion can be regarded as static in a very short time, the use of short window subtraction will produce very fuzzy results.

\subsection{Gesture recognition and physiological sign monitoring }
\label{Gesture recognition and physiological sign monitoring }
\subsubsection{Gesture recognition}
\label{Gesture recognition}
Gestures can be divided into micro gestures and macro gestures. Micro gestures generally refer to the small actions generated by a part of the hand. Macro gestures refer to the large actions of the whole hand moving from one position to another. People's research on gestures mainly focuses on the field of single hand micro movement, and the gesture recognition of simultaneous movement of both hands remains to be explored. With the penetration of computer technology in all aspects of life, human-computer interaction has become an inevitable trend. Because radar has the ability of detecting micro motion with high precision, the use of radar and other RF sensors to develop human-computer interaction interface (HCI) based on gesture recognition (HGR) has attracted the attention of researchers. In the past decade, radar based HGR has witnessed a great climax and rapid development. From the type of radar, gesture recognition can be divided into ultra bandwidth pulse radar\cite{arbabian201394,heunisch2019millimeter,park2016ir,kim2017hand} and continuous wave radar\cite{sun2020real,lien2016soli,skaria2019hand,miller2020radsense,pramudita2020time}. For pulse radar, it mainly uses the original data-driven deep learning method to recognize gestures, and the related feature extraction work is less. There are various feature extraction technologies in continuous radar. Appropriate classifiers are designed to recognize gestures through manually extracted classification features. These classifiers are based on supervised and unsupervised machine learning.

The main challenge of gesture recognition is the extraction of gesture features. Good gesture features can greatly improve the accuracy of recognition. Select appropriate radar technology for gesture motion acquisition, and then filter and format the received signal data to convert the data into appropriate format. The data format for gesture feature extraction\cite{ahmed2021hand} is mainly divided into the following six types:
\begin{itemize}
    \item Time-Amplitude : The Time-Amplitude mainly uses the time-varying amplitude of the received signal to extract the gesture motion contour. This type of gesture signal is one-dimensional and can be used to develop a simple classifier based on signal processing\cite{gao2016barcode}. 
    \item Range-Amplitude : Range-Amplitude is to extract features by using reflection intensity at different distances\cite{yeo2016radarcat}.
    \item Time-Range : The Time-Range is to obtain one-dimensional\cite{zheng2013doppler} and two-dimensional\cite{ahmed2020hand} signals to classify gestures by using the Time-Varying Distance of the received signal.That is , the amplitude of the distance of the recorded hand over time. 
    \item Time-Doppler : Time-Doppler uses Time-Varying Doppler frequency shift to extract gesture features, and takes the change of Doppler frequency with time as the input of CNN for gesture recognition\cite{kim2016hand}. 
    \item Range-Doppler frequency / Time-Doppler velocity : Range-Doppler frequency / Time-Doppler velocity extracts gesture features by the change rate of Doppler frequency shift relative to distance\cite{sang2018micro}. 
    \item Time-Frequency : Time-Frequency is to detect gestures by observing the change of frequency with time.
\end{itemize} 
Finally, after these data formats are converted into the required shapes, they can be input into an appropriate classifier for gesture classification.

\subsubsection{Physiological sign monitoring}
\label{Physiological sign monitoring}
With the development of big data technology and the concept of healthy life, more and more people begin to care about their health. In the past few years, researchers have become more and more interested in ubiquitous health detection. By monitoring people's physiological signals, such as sleep posture\cite{yue2020bodycompass}, sleep state\cite{zhao2017learning}, respiration and heart rate\cite{yue2018extracting,liu2018non,chen2020respiration,adib2015smart}, using these information to estimate a person's physical health and emotional state\cite{zhao2016emotion}, they have made great contributions to the prevention of many major diseases, It has had a great impact on the medical system. At present, many monitoring methods are either based on health monitoring equipment and need to be installed on the human body, such as finger pulse oximeter, respiratory belt, smart watch, which brings inconvenience to people's daily life. Or based on vision, this monitoring scheme not only has strong invasiveness and destroys people's privacy, but also has strong requirements for the environment, which is basically ineffective in weak light and dark environment. RF based system is welcomed by researchers because of its non-contact, low invasiveness, strong resistance to environmental factors and sensitivity to micro movement. It is widely used in the monitoring field of daily physiological signs, such as respiratory and heartbeat rate monitoring, sleep quality and state, etc.

The monitoring of respiration and heartbeat rate is based on the RF signal changes caused by small movements generated by inhalation, exhalation and heartbeat. The changes caused by such movements are regular and periodic, but this change is far less than the changes caused by static background and dynamic background in the environment. How to separate the target signal Noise elimination is a major challenge in this application field. 
In \cite{liu2018non}, the multi antenna FMCW radar is used to monitor the breathing and heartbeat rate of a single person at the same time. The unnecessary motion and interference noise in the background are reduced through the superposition of multi antenna signals and the breathing and heart rate are detected through frequency filtering and the elimination of aperiodic signals. In \cite{yue2018extracting}, the interference caused by multiple RF signals is modeled, the original respiratory signals are separated by independent component analysis (ICA), and then the respiratory signals are optimally matched with the corresponding people to realize the monitoring of multi person respiration and heart rate. In \cite{adib2015smart}, the system first locates each user in the environment, then amplifies the reflected signal of each user and analyzes its changes to extract respiration and heart rate. \cite{chen2020respiration} cross ambiguity function (CAF) and depth transmission network [DTN] When the RF signal acts on the human body, it will be reflected from the human body according to the direction of the human body to form a multi-path signal indicating the posture of the human body. By separating the signals of different paths, these signals are combined with the received signals Associate with the subjects and construct an inverse mapping to predict the sleeping position of the human body \cite{yue2020bodycompass}. \cite{zhao2017learning} combine convolution network and recursive neural network to extract the subjective change characteristics of sleep and obtain the sleep process time to predict the sleep stage.

\subsection{Human activity recognition}
\label{Human activity recognition}
Human activity recognition is not only an important branch in the field of computer vision, but also a hot spot in the research of computer vision. It is widely used in the fields of game entertainment, smart home, human-computer interaction, security and so on. Simple activity recognition is action classification. Given a section of data signal, it is correctly classified into several known action models. The activity recognition of complex points is behavior recognition. The data signal contains not only one action category, but multiple action categories. First identify the start time and end time of the action, and then classify the data action. The ultimate goal of activity recognition is to analyze who is at what time, where and what is doing in the signal. Human activity recognition can be simply divided into vision based and RF based from the form of input data.

Vision based Human activity recognition is a very extensive branch of activity recognition research at this stage. There have been many excellent methods and frameworks. These methods and frameworks can be simply divided into template based methods\cite{luo2017unsupervised,klaser2008spatio}, spatiotemporal attention based methods\cite{nazir2018evaluating,peng2016bag}, trajectory based methods\cite{wang2013action,wang2013dense}, double convolution flow based methods\cite{simonyan2014two,simonyan2015two} and three-dimensional convolution based methods\cite{tran2015learning,zhang2018rgb}. The template based method\cite{luo2017unsupervised,klaser2008spatio} uses a group of templates to represent the actions to be identified, and classifies the actions to be tested by calculating the similarity between the actions to be tested and the template actions. The limitation of this method is the definition of standard template actions, which is only suitable for classifying simple actions. The method based on spatiotemporal attention\cite{nazir2018evaluating,peng2016bag} uses the methods of detection\cite{zhang2021category,meng2021hierarchical,qiu2019a2rmnet,chen2021bal,li2019headnet} and segmentation\cite{xu2020new,shi2020query,shang2019instance,yang2020mono,yang2020learning} to extract the features of the region of interest in the video image, and send the extracted features to the classifier for classification. This method requires high accuracy of detection and segmentation, which is suitable for the situation with complex background. The idea of trajectory based method\cite{wang2013action,wang2013dense} is that a trajectory will be generated with human movement, which can be expressed by the changes of the coordinates of the key points of the human skeleton, and the actions are classified by extracting the eigenvalues of this trajectory. The method based on dual stream network convolution\cite{simonyan2014two,simonyan2015two} extracts the features of the image through two independent streams - spatial convolution stream and temporal convolution stream. Finally, the features are fused and the actions are classified. This method is not suitable for the extraction of complex optical flow information and has a large amount of calculation. Three dimensional convolution\cite{tran2015learning,zhang2018rgb} expands one-dimensional acquisition time information on the basis of two-dimensional convolution. This method is simpler than that based on two stream convolution network. It can capture short-time time characteristics more directly, but it is not suitable for capturing long-time time information.

This vision based human activity recognition is vulnerable to the influence of the environment, will completely lose its effect in the weak light or dark environment, can not work, and is vulnerable to the influence of human appearance, clothing and other characteristics. Therefore, the human activity behavior based on RF has attracted the attention of researchers. The high-frequency RF signal has a strong ability to perceive the environment and human activities, and has a strong ability to resist environmental interference. The collected RF signals are preprocessed, and then the corresponding features are extracted for action classification.

\subsubsection{Identity recognition}
\label{Identity recognition}
Identity recognition\cite{kim2014human,vandersmissen2018indoor,hsu2017extracting,hsu2019enabling,fan2020learning,garreau2011gait,tahmoush2009radar,kalgaonkar2007acoustic,zhang2008human} is an important application field in RF applications. Identity recognition is a prerequisite for multi person activity recognition. Target identity is recognized by extracting the physical characteristics of the target, which is generally used to identify people and people or human and non-human targets. The continuous wave radar is used to collect the target signal data. After eliminating the irrelevant interference and filtering and denoising, the signal in the time domain is transformed into spectrum through FFT\cite{carr1981digital}, and the features related to the physical features of the target, such as the frequency, amplitude, step, signal bandwidth and signal intensity distribution of the spectrum, are extracted from the spectrum to predict the target identity. \cite{garreau2011gait} considered MD signature to identify individuals, \cite{tahmoush2009radar} reported the results of identifying 8 individuals using a K-NN classifier and two manual features (subject's stride and trunk line). In \cite{kalgaonkar2007acoustic}, Gaussian mixture models (GMMs) were used to identify individuals and distinguish male and female subjects according to manual characteristics. A total of 20 recordings of 30 subjects were used to train and test the developed model. Similarly, in \cite{zhang2008human}, eight individuals were identified using GMM.\cite{kim2014human} SVM is trained by extracting the physical features of the target, which is used to classify the target, but it can not recognize the people who move aperiodically or fast. \cite{vandersmissen2018indoor} MD signature is used to represent the two-dimensional spatial structure. As the input of DCNN, gait features are extracted to identify people moving freely in space, which makes the model more robust to deal with environmental changes. \cite{fan2020learning} Taking the RF track as the input, a hierarchical attention module is used to generate the feature map, and the multi-task learning method of recognizable features is used as supervision to extract a long-term recognition feature to recognize the target identity.

\subsubsection{Activity recognition}
\label{Activity recognition}
\textbf{Action recognition: }
In the traditional behavior recognition model, the features are usually extracted manually\cite{avrahami2018below,amin2020radar,qi2019position,zhu2018indoor,alnaeb2019detection,shrestha2020continuous,chetty2017low}, and then classified by classifier.In\cite{zhu2018indoor}, a human motion recognition framework based on environmental radar is proposed. A high-frequency radar is used to sample the reflected signal to capture the fine state of human activities. It has great limitations to classify some simple actions according to the manually extracted slope and relative position features. In\cite{avrahami2018below} , a method of using RF data projection splicing to improve the accuracy of behavior recognition is proposed and tested in two environments. In\cite{amin2020radar} solved the problem of continuity of human movement and used the nature of human activities to eliminate some impossible movements. In this study, the action is the conversion process of two states. The behavior can be described by the change from one state to another. The micro Doppler features and target distance extracted from the radar signal spectrum are input into two-dimensional PCA, and then classified by K-NN classifier. \cite{qi2019position} used the Super bandwidth biological radar to classify human activities based on Doppler signals, and proposed a location information index classifier (PIIC), which improved the performance of the classifier. In \cite{guendel2020derivative}, a DTL based method is proposed to separate motion activities from in-situ activities, then 2D PCA is used to extract features, and decision time classifier is used to classify actions. \cite{alnaeb2019detection} used STFT to extract spectral features and SVM to detect falls.

\textbf{Behavior recognition: }
With the increasing amount of data and the complexity of experimental environment, the limitations of traditional behavior recognition patterns are becoming more and more prominent. In the past few years, the progress of deep learning\cite{hsu2019enabling,vasisht2018duet,uddin2020home,li2019making,guo2020deep,fan2020home} has promoted the progress of behavior recognition at an amazing speed. Therefore, researchers use deep learning to extract more advanced features to identify human behavior, and many excellent behavior recognition models based on deep learning have emerged, such as RF action\cite{li2019making}, duet\cite{vasisht2018duet}, Marko\cite{hsu2019enabling}. \cite{guo2018real} describes a data acquisition method, which generates a series of three-dimensional images according to the signal intensity reflected by the human body, inputs this continuous signal image into CNN in \cite{guo2020deep}, extracts the spatial feature sequence, and extracts the temporal feature into RNN for human behavior recognition. \cite{shrestha2020continuous} input radar data in the form of continuous time series of Doppler or range time information into Bi-LSTM or LSTM, and analyze continuous human activity sequences by using the time dependence between samples. \cite{uddin2020home} A behavior situation classification model based on RNN is proposed. In \cite{li2019making}, a neural network model for detecting human behavior through the wall is proposed, and the action recognition model based on RF skeleton is used to understand people's behavior and interaction. The RF signal is used as the input of the network to generate 3D human bones, which are input into the spatiotemporal attention module to extract the features of people of interest, and then these features are transmitted and input into the multi proposal to identify multi person actions and interactive actions. \cite{hsu2017extracting} Based on the progress of passive wireless location in the past, by continuously tracking people's trajectory information to study family social behavior, an algorithm combining time and space is designed to generate human trajectory. The interaction behavior in three situations is analyzed in the form of trajectory + RF signal. \cite{hsu2019enabling} A multi-modal system is designed to infer the location and behavior of users through four components - identity matching subsystem, HMM, first-order logic framework and radio localization subsystem. Firstly, the trajectory is generated according to the RF signal, the trajectory is used as the input of HMM, and the identity of the trajectory data is matched according to the intermittent interaction between the user and the mobile phone to output the probability. The first-order logic component obtains the output of identity matching and HMM, and generates the best user behavior explanation. \cite{fan2020home} proposed a framework for generating caption description behavior using RF signal, which is composed of four modules. The first module encodes and embeds RF skeleton and floormap together as the feature of the first module. The second module uses the video encoder to encode the paired and unpaired video after pooling as the feature of the second module. The third module aligns the video features and RF features to form a unified feature spectrum. The fourth module is the language generation module, which inputs the unified feature spectrum into the language generation network to generate words to describe human behavior.

\subsection{Human pose estimation}
\label{Human pose estimation}
\textbf{Voxel power: }
\cite{adib2015capturing} introduces a method of capturing human body contour to generate skeleton, RF capture, which scans 3D space by using 2D antenna array and FMCW chirps, first performs coarse resolution scanning to obtain the area with high reflection power, and then performs fine resolution scanning to amplify the area with reflection power, so as to obtain RF frames. The coordinates of voxels in each 3D space are represented by spherical coordinates $(r , \theta , \phi)$, The power of each voxel in the frame can be calculated by the following formula:
\begin{equation}
    \label{eq11}
    P(r , \theta , \phi) = \left\lvert \sum_{m = 1}^{M}\sum_{n = 1}^{N}\sum_{t = 1}^{T}S_{m,n,t}e^{j2\pi\frac{kr}{c}t}e^{j\frac{2\pi}{\lambda}\sin\theta(nd\cos\phi + md\sin\phi)}  \right\rvert 
\end{equation}
In this formula, $M , N$ represents the number of transmitting antennas and the number of receiving antennas respectively, $S_{m,n,t}$ represents the signal from the transmitting antenna $m$ to the receiving antenna $n$ at time $t$.

\textbf{Pose estimation: }
RF based pose estimation faces a huge challenge -- there is no large amount of annotation data. In order to train a reliable network to estimate human pose, we need to use the idea of cross-modal supervision. As an intermediate representation feature, skeleton provides the basis for cross-modal supervision. The human body is an irregular body. In the process of human movement, some RF signals will not be directly reflected to the receiving antenna. The human body information in a single RF frame is incomplete and only contains some limb information. This is the first challenge for RF skeleton generation. \cite{adib2015capturing} Through the cross frame joint method, the complete human body information can be obtained, and then a coarse-grained 2D RF-skeleton can be obtained through depth compensation, swing compensation, human body segmentation and bone suture. 
\cite{zhao2018through} is a device free attitude estimation system RF-pose, which does not need to wear any sensors to extract accurate two-dimensional human posture by analyzing RF signals. The biggest problem of using RF to estimate human posture is that there is no labeled data, so it is impossible to generate a two-dimensional skeleton with high reliability. In this study, a cross-modal supervisory network model Teacher-Student is proposed to solve the problem of no labeled data. Synchronize RF and video streams by connecting a camera to a wireless sensor. The video stream is used as the input of the Teacher network and the RF stream is used as the input of the Student network. The information of human joint points is extracted from the video stream as the supervised training Student network to generate fine-grained 2D RF skeleton. After training, attitude estimation can be performed using only RF signals as input. In \cite{zhao2018rf}, human posture is estimated by generating 3D RF skeleton. The RF signal input by the system is a 4D tensor, which will greatly increase the amount of calculation and can not dynamically display the skeleton action in real time. Therefore, RF-pose3D projects this 4D tensor to different planes and decomposes it into two 3D tensors, using horizontal RF frames and vertical RF frames as the input of  RF network, The computation of convolution network is greatly reduced. A multi-camera coordination system is designed to generate a video based 3D skeleton to train the 3D RF skeleton for estimating human posture. Based on the real-time 3D human posture estimation system with equipment\cite{yang2020subject, yang2020rfid}, the human posture is constructed by mapping the position of the additional RFID tag to 3D coordinates. Since the original RFID phase data will be seriously distorted due to channel jump and phase winding, first calibrate the phase, and use high-precision low rank tensor completion (HaLRTC) to interpolate the missing RFID data, and then conduct down sampling operation to synchronize with the 3D pose time series obtained by Kinect 2.0, so as to ensure the high accuracy and real-time performance of pose estimation, Finally, the calibrated phase data is used as input to train the depth neural network of human skeleton reconstruction, and the labeled visual data collected by Kinect 2.0 device is used as supervision.

\section{Conclusion}
\label{Conclusion}
With the popularity of security awareness, people pay more and more attention to their privacy. The environment sensing technology based on wireless signal has attracted extensive attention and attention of scholars because of its unique advantages - no need to wear any sensor, non line of sight sensing, less affected by external conditions, strong scalability, strong privacy protection and so on. It can be seen from the above articles that researchers try to start from two directions and hope to design and build an environment independent, location independent and life-oriented RF based environment perception technology in the future. One is the development direction of hardware, which designs an ultra-high frequency and ultra-bandwidth device to collect signals with richer characteristics. The other is the direction of software development. By considering more influencing factors, we design an ultra-high-performance algorithm to eliminate interference and apply it in all aspects of life.

\bibliography{main}
\bibliographystyle{IEEEtran}

\end{document}